\documentstyle[12pt]{article}

\author{Chi Au and Bao-Heng Zhao\thanks{Permanent address: Department of Physics, Graduate School, Chinese Academy of Scienses, P. O. Box 3908, Beijing 100039, China}\\
Department of Applied Mathematics\\
The Hong Kong Polytechnic University, Hong Kong
\and
H. T. Nieh\thanks{On leave of absence from Institute for Theoretical Physics, State University of New York at Stony Brook, Stony Brook, NY 11794, USA}\\
The Hong Kong University of Science and Technology, Hong Kong}
\title{Phase Separation in the Hubbard Model
}
\date{%
}

\begin{document}

\maketitle
\begin{abstract}
Making use of symmetry properties and known exact results for the Hubbard
Model on bipartite lattice, we show that (1) there is no phase separation
for repulsive coupling at low dopings, (2) phase separation and
superconductivity co-exist in the ground states for attractive coupling in a
range of filling fractions.

\end{abstract}

\newpage\ Recent experiments indicate the presence of phase separation in
cuprate superconductors, namely, there exist separate hole-rich and
hole-poor phases\cite{1}-\cite{4}. There have been theoretical studies, all
employing approximations or numerical estimations, on this phenomenon based
on the {\it t-J} model\cite{5}-\cite{6}, as well as directly on the Hubbard
model\cite{7}-\cite{8}. Most of these studies indicate that these models can
explain the phenomenon of phase separation. The exceptions are those of
Putikka {\it et al.}\cite{6}, and Moreo {\it et al}. \cite{7}. While the
quantum Monte Carlo calculations of small 2D lattice systems indicate no
phase separation for the 2D Hubbard model\cite{7}, the investigation using
high-temperature expansions for the 2D {\it t-J} model suggests the
existence of phase separation only for a limited range of {\it J/t} values%
\cite{6}.

In this note, we look again at this phenomenon on the basis of the Hubbard
model with repulsive on-site interaction. We make use of some exact results
for the Hubbard model on bipartite lattice with attractive coupling\cite{9}-%
\cite{10}, which, after applying the well-known unitary particle-hole
transformation\cite{11}, imply a certain general property for the repulsive
coupling case. It in turn implies that the repulsive Hubbard model on
bipartite lattice can not accommodate phase separation at low dopings. We
also look at the case of attractive Hubbard model, for which we show that,
at temperature $T=0,$ phase separation is possible, and actually co-exists
with superconductivity.

The Hamiltonian of the Hubbard model on bipartite lattice is 
\begin{equation}
\label{1}H=-\sum\limits_{ij,\sigma }t_{ij}c_{i\sigma }^{\dagger }c_{j\sigma
}+U\sum\limits_ic_{i\uparrow }^{\dagger }c_{i\uparrow }c_{i\downarrow
}^{\dagger }c_{i\downarrow }-\mu \sum_i(c_{i\uparrow }^{\dagger
}c_{i\uparrow }+c_{i\downarrow }^{\dagger }c_{i\downarrow }), 
\end{equation}
with the usual notations. The bipartite lattice sites $\Lambda $ are divided
into subsets {\it A} and {\it B. }The number of sites of {\it A (B)} is
denoted by $\left| A\right| $ ($\left| B\right| $), and the total number of
sites by $\left| \Lambda \right| .$ In (1), $t_{ij}$ vanishes if both {\it i 
}and {\it j} belong to the same subset. An exact result obtained by Lieb\cite
{9} states that, for attractive coupling ($U<0$) and even number of
electrons, the ground state of the system is unique and has total spin $S=0$%
. Thus, at temperature $T=0,$ the system is non-ferromagnetic. This result
was extended to $T>0$ by Kubo and Kishi\cite{10}, who find that there is an
upper bound on a certain two-point function. It implies absence of
long-range magnetic order in attractive Hubbard model at finite temperature.

We introduce an external magnetic field in the {\it z}-direction. The
Hamiltonian (1) becomes 
\begin{equation}
\label{2} 
\begin{array}{c}
H=-\sum\limits_{ij,\sigma }t_{ij}c_{i\sigma }^{\dagger }c_{j\sigma
}+U\sum\limits_ic_{i\uparrow }^{\dagger }c_{i\uparrow }c_{i\downarrow
}^{\dagger }c_{i\downarrow }-\mu \sum_i(c_{i\uparrow }^{\dagger
}c_{i\uparrow }+c_{i\downarrow }^{\dagger }c_{i\downarrow }) \\ 
+h\sum_i(c_{i\uparrow }^{\dagger }c_{i\uparrow }-c_{i\downarrow }^{\dagger
}c_{i\downarrow }). 
\end{array}
\end{equation}
Define 
\begin{equation}
\label{3}s^z=\frac{\sum_is_i^z}{\left| \Lambda \right| }, 
\end{equation}
where $s_i^z$ is the {\it z}-component of the spin of the particle on site $%
i $. In general, we have 
\begin{equation}
\label{4}\lim _{\left| \Lambda \right| \rightarrow \infty }\left\langle
s^z\right\rangle =f(h,T), 
\end{equation}
where $\left\langle ...\right\rangle $ is the average over a grand canonical
ensemble. Invoking Lieb-Kubo-Kishi's result, we have 
\begin{equation}
\label{5}f(0,T)=0, 
\end{equation}
and, furthermore, since there is no long-range magnetic order, that $f(h,T)$
is analytic at $h=0$. As a consequence, we can draw the conclusion that {\it 
$f(h,T)$ is a well-behaved, i. e. continuous and single-valued, function in
the neighborhood of $h=0$ for $T\geq 0.$} For small $h,$%
\begin{equation}
\label{6}\lim _{\left| \Lambda \right| \rightarrow \infty }\left\langle
s^z\right\rangle =\chi (T)h, 
\end{equation}
where $\chi (T)$ is the spin susceptibility.

We now make the unitary particle-hole transformation\cite{11} 
\begin{equation}
\label{7}c_{i\uparrow }\rightarrow c_{i\uparrow },\,\,c_{i\downarrow
}\rightarrow \epsilon _ic_{i\downarrow }^{\dagger }, 
\end{equation}
where $\epsilon _i=1$ ($-1$) on the lattice set $A$ ($B$). The Hamiltonian
(2), up to a constant, is transformed into 
\begin{equation}
\label{8} 
\begin{array}{c}
\tilde H=-\sum\limits_{\left\langle ij\right\rangle ,\sigma
}t_{ij}c_{i\sigma }^{\dagger }c_{j\sigma }+U^{\prime
}\sum\limits_ic_{i\uparrow }^{\dagger }c_{i\uparrow }c_{i\downarrow
}^{\dagger }c_{i\downarrow }-\mu ^{\prime }\sum_i(c_{i\uparrow }^{\dagger
}c_{i\uparrow }+c_{i\downarrow }^{\dagger }c_{i\downarrow }) \\ 
+h^{\prime }\sum_i(c_{i\uparrow }^{\dagger }c_{i\uparrow }-c_{i\downarrow
}^{\dagger }c_{i\downarrow }), 
\end{array}
\end{equation}
with 
\begin{equation}
\label{9} 
\begin{array}{c}
U^{\prime }=-U, \\ 
\,\,\, 
\frac{U^{\prime }}2-\mu ^{\prime }=h, \\ \,\,\,h^{\prime }=\frac U2-\mu , 
\end{array}
\end{equation}
and (6) into 
\begin{equation}
\label{10}\lim _{\left| \Lambda \right| \rightarrow \infty }D=1+\chi
(T)(U^{\prime }-2\mu ^{\prime }), 
\end{equation}
where $D$ is the density for the system described by $\tilde H,$ 
\begin{equation}
\label{11}D=\frac{\sum_{i,\sigma }\left\langle c_{i\sigma }^{\dagger
}c_{i\sigma }\right\rangle }{\left| \Lambda \right| }. 
\end{equation}
Recall that $U^{\prime }=-U>0.$

From (10), we see that for a system described by a repulsive Hubbard model ($%
U^{\prime }>0$) on bipartite lattice, the density $D$ in the thermodynamic
limit is a well-behaved, i. e.{\it \ continuous} {\it and single-valued},
function of the chemical potential $\mu ^{\prime }$ near $\frac{U^{\prime }}%
2 $, namely at low dopings. This behavior is in contrast with what is
expected of a first-order phase transition.

When a system with chemical potential $\mu _o$ undergoes a first-order phase
transition at temperature {\it T}, and separates into two phases with
different densities, we have 
\begin{equation}
\label{12} 
\begin{array}{c}
D_1=\lim _{\mu \rightarrow \mu _o+0}\lim _{\left| \Lambda \right|
\rightarrow \infty }D, \\ 
D_2=\lim _{\mu \rightarrow \mu _o-0}\lim _{\left| \Lambda \right|
\rightarrow \infty }D, 
\end{array}
\end{equation}
with $D_1\neq D_2.$ The density {\it D}, as a function of the chemical
potential, is discontinuous at $\mu _o$ in the thermodynamic limit when
there is first-order phase transition. Thus the continuity of $\lim _{\left|
\Lambda \right| \rightarrow \infty }D$ in (10) implies that {\it a repulsive
Hubbard model on bipartite lattice can not accommodate co-existence of two
phases at low dopings. }This result is valid for general $h^{\prime },$
including $h^{\prime }=0,$ which is the case of vanishing external magnetic
field.

The following are two remarks:

1.The conclusion we made with regard to the phase separation in a repulsive
Hubbard model on bipartite lattice applies to low dopings. This is because
the exact result of Lieb-Kubo-Kishi is for the case of vanishing magnetic
field ($h=0$). However, if the function $f(h,T)$ is also well behaved, as
may be reasonably expected for a paramagnetic system, at large values of 
{\it h}, then our conclusion about phase separation can be extended to high
dopings.

2. Lieb has proved\cite{9} in the case of repulsive Hubbard model on
bipartite lattice with half-filling and $h=0$ that the (degenerate) ground
states have the same spin $S=\frac 12(\left| A\right| -\left| B\right| )$
(if $\left| A\right| \geq \left| B\right| $), with 
\begin{equation}
\label{13}S_z=-\frac 12(\left| A\right| -\left| B\right| ),\,\,-\frac
12(\left| A\right| -\left| B\right| )+1,\,\,...,\frac 12(\left| A\right|
-\left| B\right| ) 
\end{equation}
This result can be used to make certain statement concerning phase
separation at $T=0$ for attractive Hubbard model on bipartite lattice.
Making use again of the unitary particle-hole transformation, it is
straightforward to obtain, for the attractive case, that at $T=0,\,$ 
\begin{equation}
\label{14}D=1-\frac{\left| A\right| -\left| B\right| }{2\left| \Lambda
\right| },\,\,1-(\frac{\left| A\right| -\left| B\right| }{2\left| \Lambda
\right| }-\frac 1{\left| \Lambda \right| }),\,\,...,1+\frac{\left| A\right|
-\left| B\right| }{2\left| \Lambda \right| }. 
\end{equation}
This implies that in an attractive Hubbard model on bipartite lattice
different phases can co-exist at $T=0,$ with the densities having values
given in (14).

It has been shown by Shen and Qiu\cite{12} that for attractive Hubbard model
on bipartite lattice with {\it $\left| A\right| -\left| B\right| =O(\left|
\Lambda \right| ),$} the ground states with densities in the range {\it $(1- 
\frac{\left| A\right| -\left| B\right| }{2\left| \Lambda \right| })$ $<D<(1+ 
\frac{\left| A\right| -\left| B\right| }{2\left| \Lambda \right| })$} have
off-diagonal long-range order (ODLRO) in the two-particle reduced density
matrix, and, consequently, posses the property of superconductivity\cite{13}%
. That is, superconductivity and phase separation co-exist in this case.

In conclusion, we have investigated the question of phase separation for the
Hubbard model on bipartite lattice. Our result that there is no phase
separation at low dopings for $T\geq 0$, being based on the exact results of
Lieb\cite{9}, and Kubo-Kishi\cite{10}, is an {\it exact} statement of the
repulsive Hubbard model on bipartite lattice of any dimension. It is
consistent with the quantum Monte Carlo calculations of small 2D lattice
systems for the Hubbard Model\cite{7}. It is also consistent with the
high-temperature expansion result of Putikka {\it et al.}\cite{6} for the 2D 
{\it t-J} model, to the extent that the Hubbard model and the {\it t-J}
model are expected to have similar results at small {\it J/t} values.

\noindent {\bf Acknowledgment}

We acknowledge discussions with Prof. Fu-Chun Zhang, and thank him for
pointing out the work of Putikka et al.\cite{6} This work is supported in
part by The Hong Kong Polytechnic University. One of the authors (B. H.
Zhao) is grateful to the hospitality of the Department of Applied
Mathematics, The Hong Kong Polytechnic University. He is also supported in
part by NNSF of China.

\end{document}